# Wave propagation in tunable lightweight tensegrity metastructure

Y.T. Wang, X.N. Liu, R. Zhu[*], G.K. Hu[*]

*Key Laboratory of Dynamics and Control of Flight Vehicle, Ministry of Education, School of Aerospace Engineering, Beijing Institute of Technology, Beijing 100081, China*

## ABSTRACT

In this paper, lightweight metastructures are designed consisting of prismatic tensegrity building blocks which have excellent strength-to-weight ratio and also enable unique compression-torsion coupling. A theoretical model with coupled axial-torsional stiffness matrix is first developed to study the band structures of the proposed lightweight metastructures. Unit cell designs based on both Bragg scattering and local resonance mechanism are investigated to generate bandgaps at desired frequency ranges. Broadband full-wave attenuation is found in the tensegrity metastructure with special opposite-chirality unit cells. Furthermore, tunable stiffness in the prismatic tensegrity structure is investigated and 'small-on-large' tunability in the tensegrity metastructure is achieved by harnessing the geometrically nonlinear deformation through an external control torque. Prestress adjustment for fine tuning of the band structure is also investigated. Finally, frequency response tests on finite metastructures are preformed to validate their wave attenuation ability as well as their wave propagation tunability. The proposed tensegrity metastructures could be very useful in various engineering applications where lightweight and tunable structures with broadband vibration suspension and wave attenuation ability are in high demand.

*Email: rzhu83ac@gmail.com
*Email: hugeng@bit.edu.cn

## 1. Introduction

Tensegrity structures are lightweight spatial structures with a highly efficient material utilization and therefore, can form minimal mass system with load-bearing capability (Masic, et al., 2006; Juan and Tur, 2008; Feng, et al., 2010; Skelton, et al., 2014; Skelton, et al., 2016) . Typically, tensegrity structures consist solely of bars and strings and own their shapes and stiffness to the tensile stress in the strings. As a result, the mechanical response of tensegrity can be easily adjusted with topology of connections, masses' shapes and positions and the prestress of the strings (Shea, et al., 2002; Fest, et al., 2003; Fest, et al., 2004; Ali and Smith, 2010). Such unique properties make tensegrities very desirable in various lightweight-emphasized engineering structures and adaptive applications, such as deployable aerospace and civil structures (Fu, 2005; Fazli and Abedian, 2011; Fraternali, et al., 2015), robotics (Graells Rovira and Mirats Tur, 2009; Moored, et al., 2011; Caluwaerts, et al., 2014) and sensors/actuators (Skelton, 2002; Sultan and Skelton, 2004). Recently, tensegrity concept has also been successfully employed to describe mechanics of some biological structures (Ingber, 1997; Wang, et al., 2001; Liu, et al., 2004; Luo, et al., 2008).

Prismatic tensegrity structure (PTS) represents one of the simplest forms of tensegrities and its geometrical and elastic behaviors have been well studied in the past (Oppenheim and Williams, 2001; Schenk, et al., 2007; Zhang, et al., 2009; Li, et al., 2010; Zhang, et al., 2014; Zhang and Xu, 2015; Ashwear, et al., 2016; Zhang, et al., 2016; Cai, et al., 2017) . Most recently, novel PTS's properties regarding the extreme mechanical responses as well as the nonlinear wave propagations have been discovered. Statically, Oppenheim and Williams (2000) studied PTS models which demonstrated extreme stiffening-type response in the presence of rigid bases. Amendola et al. (2014) developed new assembly methods of bi-material PTS and experimentally investigated its compressive response in the large displacement regime where switches from stiffening response to softening response were discovered. Fraternali, et al. (2015) studied the geometrically nonlinear behavior of uniformly compressed PTS through full elastic and rigid-elastic models and both extreme stiffening and softening behaviors were observed. Dynamically, one-dimensional (1D)

PTS array has been explored as a waveguide to support energy transportation through solitary waves (Fraternali, et al., 2012). Later on, Fraternali, et al. (2014) utilized the softening and hardening regimes of a PTS chain to tune solitary rarefaction and compression waves which exhibit anomalous wave transmission and reflection. By gradually changing PTS's elastic response from stiffening to softening through the modification of mechanical, geometrical, and prestress variables, solitary waves with designable wave profile can also be achieved and potential applications such as tunable acoustic lenses were demonstrated (Fabbrocino and Carpentieri, 2017). More interestingly, the naturally coupled axial and torsional motions in a chiral-shape PTS can be explored to design special micropolar materials (Liu, et al., 2012) that go beyond Cauchy continuum mechanics and provide peculiar static material behaviors (Frenzel, et al., 2017). For wave dynamics, PTSs with rich and potentially controllable elastic behaviors can be excellent building blocks to form phononic crystals or locally resonant (LR) elastic metamaterials for simultaneously lightweight and functional wave material systems.

Metastructure, as a metamaterial inspired concept, has recently emerged to refer to a structure-like periodic material system with excellent wave absorption abilities and stiffness-to-weight ratio (Hussein, et al., 2014; Reichl and Inman, 2017). Hussein, et al., (2014) introduces metastructure as an emerging research field involving vibration/acoustic engineering and condense matter physics. Reichl and Inman (2017) emphasized on the lightweight and vibration absorption abilities of the metastructure with optimized microstructure geometries. Although tailoring the geometric and elastic properties of the metastructure's building blocks could tune its wave behavior (Liu, et al., 2011; Zhu, et al., 2011), a broadband design still requires additional unit cells (Zhu, et al., 2014) which inevitably increase the overall weight of the engineering structure. Moreover, like any passive metamaterials or phononic crystals, once the unit cell is manufactured, changing the position and size of metastructure's bandgap would be very difficult in practice, if not impossible. One good solution for actively controlling the wave behavior of the metastructure is to introduce electromechanical coupling which provides an externally controllable degree of freedom in each unit cell (Airoldi and Ruzzene, 2011; Deue, et al., 2014; Wang, et al., 2016; Chen, et al., 2016). Zhu et al. (2016) fabricated an adaptive metastructure with plastic tube and beam elements with surface-bonded piezoelectric patches and demonstrated that its bandgaps can be fully tailored by adjusting parameters of the shunted electric circuits. With the help of hardening and softening shunted circuits, tunable bandgap capacity as high as 45% was achieved experimentally. However, it was also observed during the experiment that each shunted circuit requires independent adjustment due to the unavoidable inconsistency among manufactured metastructure's unit cells, which could bring difficult in practical applications. The complicated stability condition in the control circuits could also become a problem to the robustness of the active metastructure (Zhu, et al., 2016). An alternative solution to achieve tunable metastructure can be found without coupling with the other physical fields, which could significantly promote manufacturing feasibility of the unit cell as well as decrease the complexity of the entire system. Utilizing nonlinear elastic deformations, control of the small-amplitude linear wave in phononic crystals (Wang, et al., 2013) as well as LR-based elastic metamaterial (Wang, et al., 2014) have been demonstrated. Naturally, one can expect interesting and practically tunable elastic wave functions in PTS-based metastructure where geometric nonlinearity and compression-torsion coupling can be found intrinsically in those lightweight structures.

In this paper, a theoretical model is developed to investigate the unique compression-torsion coupling in a PTS unit cell through an effective stiffness matrix. Tunable stiffness and dispersion curves of the periodically-ranged PTSs are observed under a torque-induced nonlinear deformation. Furthermore, lightweight metastructure designs based on both Bragg scattering and local resonance mechanism are investigated for different targeted frequency ranges. Broadband isolation for axial and torsional vibrations is observed in a tensegrity metastructure with unit cells having opposite chirality. Tunable wave propagations are achieved in the proposed tensegrity metastructures by two approaches: (i) harnessing the geometrically nonlinear deformation of the PTSs under global control torque; (ii) adjusting the prestress in the tension strings for small-range and fine adjustment of the band structure. Finally, frequency responses of the finite metastructures under different loadings are numerically investigated to validate the band structure results.

## 2. Theoretical model of tunable prismatic tensegrity structure

A schematic of the studied PTS is shown in Fig. 1a, which is modified from the well-studied T3 module (Oppenheim and Williams, 2000). The PTS consists two Aluminum (Al) disks at the top and bottom ends, which are then connected with three Nylon cross-strings (gray colored) and three polylactic acid (PLA) bars (yellow colored) in a right-handed chiral fashion. Due to the large differences in the material properties between the disks and the string/bars, the mass of the strings/bars can be ignored and the disks can be considered rigid in the following study. The gray spheres in Fig. 1a represent the spherical joints that permit rotational degrees of freedom (DOFs) of the bars and strings. A reference configuration of the PTS is shown in Fig. 1b, where the radius of the end-disks, the height of the PTS and the relative angle of the two end-disks are $R$, $h$ and $\phi$, respectively. In this study, we assume that the two end-disks are maintained to be parallel and the central axis of the PTS, $OO'$, is always along the z direction. Therefore, only two DOFs of the PTS are permitted, which are the relative rotational angle and the relative axial displacement between the two end-disks.

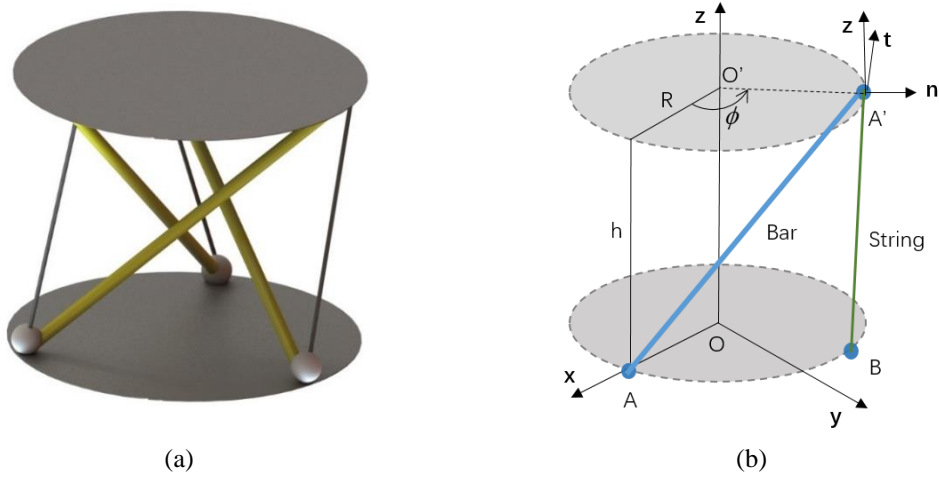

(a)                  (b)
**Fig. 1.** (a) Schemitc of the PTS; (b) reference configuration of the PTS.

The equilibrium equation at joint $A'$ can then be expressed by using a local coordinate system $n$-$t$-$z$ as:

$$p_s \frac{(\mathbf{OA'} - \mathbf{OB})}{|\mathbf{OA'} - \mathbf{OB}|} + p_b \frac{(\mathbf{OA'} - \mathbf{OA})}{|\mathbf{OA'} - \mathbf{OA}|} + p_d \frac{(\mathbf{OA'} - \mathbf{OO'})}{|\mathbf{OA'} - \mathbf{OO'}|} = \mathbf{f} \qquad (1)$$

where, $\mathbf{f} = (f_n, f_t, f_z)$ is the force applied at A′, $p_s$, $p_b$ and $p_d$ are the magnitudes of the forces along the string, the bar and the radius of the disk plane, respectively. The unit direction vectors in Eq. (1) are defined as

$$\frac{\mathbf{OA'} - \mathbf{OA}}{|\mathbf{OA'} - \mathbf{OA}|} = \frac{[R-R\cos(\phi)]}{\sqrt{R^2[2-2\cos(\phi)]+h^2}} \mathbf{n} + \frac{R\sin(\phi)}{\sqrt{R^2[2-2\cos(\phi)]+h^2}} \mathbf{t} + \frac{h}{\sqrt{R^2[2-2\cos(\phi)]+h^2}} \mathbf{z}$$

$$\frac{\mathbf{OA'} - \mathbf{OB}}{|\mathbf{OA'} - \mathbf{OB}|} = \frac{[R-R\cos(\frac{2}{3}\pi-\phi)]}{\sqrt{R^2[2-2\cos(\frac{2}{3}\pi-\phi)]+h^2}} \mathbf{n} - \frac{R\sin(\frac{2}{3}\pi-\phi)}{\sqrt{R^2[2-2\cos(\frac{2}{3}\pi-\phi)]+h^2}} \mathbf{t} + \frac{h}{\sqrt{R^2[2-2\cos(\frac{2}{3}\pi-\phi)]+h^2}} \mathbf{z} \qquad (2)$$

$$\frac{\mathbf{OA'} - \mathbf{OO'}}{|\mathbf{OA'} - \mathbf{OO'}|} = \mathbf{n}$$

Eq. (1) can also be written in a matrix form as

$$\mathbf{CP} = \mathbf{f} \qquad (3)$$

where

$$\mathbf{C} = \begin{bmatrix} \frac{R-R\cos(\phi)}{\sqrt{R^2[2-2\cos(\phi)]+h^2}} & \frac{R\sin(\phi)}{\sqrt{R^2[2-2\cos(\phi)]+h^2}} & \frac{h}{\sqrt{R^2[2-2\cos(\phi)]+h^2}} \\ \frac{R-R\cos(\frac{2}{3}\pi-\phi)}{\sqrt{R^2[2-2\cos(\frac{2}{3}\pi-\phi)]+h^2}} & \frac{-R\sin(\frac{2}{3}\pi-\phi)}{\sqrt{R^2[2-2\cos(\frac{2}{3}\pi-\phi)]+h^2}} & \frac{h}{\sqrt{R^2[2-2\cos(\frac{2}{3}\pi-\phi)]+h^2}} \\ 1 & 0 & 0 \end{bmatrix}, \quad \mathbf{P} = \begin{Bmatrix} p_s \\ p_b \\ p_t \end{Bmatrix}, \quad \mathbf{f} = \begin{Bmatrix} f_n \\ f_t \\ f_z \end{Bmatrix}.$$

A so-called 'tensegrity configuration' can be formed when the PTS is in a stable equilibrium configuration under null nodal force (Juan and Tur, 2008). In such configuration, the homogeneous equation **CP=0** should have a nontrivial solution so that $p_s$ is positive and therefore, no string is under compression. As a result, the determinate of the matrix **C** should be zero as

$$\det(\mathbf{C}) = 0 \qquad (4)$$

which results in two possible tensegrity configurations with $\phi_0 = -\pi/6$ or $5\pi/6$. However, $\phi_0 = -\pi/6$ is an unstable equilibrium configuration since this position always yields a maximum of potential energy (Oppenheim and Williams, 2000). Therefore, the only stable tensegrity configuration requires $\phi_0 = 5\pi/6$.

For fixed bottom disk, the top disk of the PTS permits coupled axial and torsional motions which can be defined as $u$ and $\theta$, respectively. It is noticed that the value of $\theta$ should be constrained within $-7\pi/6$ and $\pi/6$ to avoid the touch of the strings and the bars, respectively (Oppenheim and Williams, 2000). For the tensegrity structure under external axial force and torque, the total potential energy of the system, $\Pi$, is the sum of structure strain energy and the work down by external force and torque as

$$\Pi = E - T\theta - Fu \tag{5}$$

where $T$ and $F$ denote the externally applied torque and force, respectively. The principle of minimum potential energy is applied to determine the equilibrium of the system. Thus, force and torque on the PTS can be expressed by the axial and rotational displacements of the PTS, whose first order Taylor expansion can be obtained as $T = \frac{\partial T}{\partial \theta}\theta + \frac{\partial T}{\partial u}u$ and $F = \frac{\partial F}{\partial \theta}\theta + \frac{\partial F}{\partial u}u$, respectively. The obtained governing equations of the PTS and the detailed derivations can be found in Appendix. Equivalently, the PTS can be represented by a homogenous elastic bar with governing equation expressed as:

$$\begin{Bmatrix} F \\ T \end{Bmatrix} = \begin{bmatrix} k_h & k_c \\ k_c & k_m \end{bmatrix} \begin{Bmatrix} u \\ \theta \end{Bmatrix} \tag{6}$$

where $k_h$, $k_c$ and $k_m$ are the effective axial stiffness, effective coupling stiffness and effective rotation stiffness, respectively. By comparing with the governing equations of the PTS, the components in the effective stiffness matrix can then be obtained (detailed expressions are in Appendix). Since geometrical nonlinear behavior can be found intrinsically in the PTS (Fraternali, et al., 2015; Fraternali, et al., 2012; Zhang, et al., 2013), large modifications in the structural configuration induced by the external loadings could consequently vary the effective stiffness matrix and therefore, change the static as well as dynamic responses of the PTS. Fig. 2a shows that each component in the obtained effective stiffness matrix can be represented by a nonlinear function of the axial and rotational displacements of the PTS. In the figure, the normalized stiffness is defined as $k_h^* = k_h/k_{h0}$, $k_c^* = k_c/k_{c0}$ and $k_m^* = k_m/k_{m0}$ with $k_{h0}$, $k_{c0}$ and $k_{m0}$ being the effective stiffness at the tensegrity configuration ($u = 0$, $\theta = 0$). For a stable structure, $k_h^*$ and $k_m^*$ should always be greater than zero (Zhang and Xu, 2015; Zhang, et al., 2016) and therefore, the deep blue regions ($k_m^* < 0$) in Fig. 2a represent unstable PTS configurations which will not be studied in this research. Adjusting the effective stiffness matrix through geometrical nonlinearity suggests an attractive approach for the in-situ tuning of a wave system constructed by the PTS cells (Bertoldi and Boyce, 2008; Wang, et al., 2013). Also, it is very interesting to notice that the effective stiffness changes more dramatically with the rotational angle than its changing with the axial displacement, which indicates that a static torque loading can be applied to the PTS as a more efficient way to adjust the PTS's effective elastic properties and furthermore, control the dynamic behavior. Fig. 2b shows the static torque adjustments on the $k_h^*$, $k_c^*$ and $k_m^*$, where normalization on the torque is applied as $T^*=T/k_{m0}$. In the figure, $k_h^*$ and $k_c^*$ decrease monotonically when $T^*$ increases while $k_m^*$ decreases when $T^*<0$ but slightly increases when $T^*>0$, which suggests different control strategies for the axial and torsional wave propagations in the PTS-based wave systems as we will explain them in the following part.

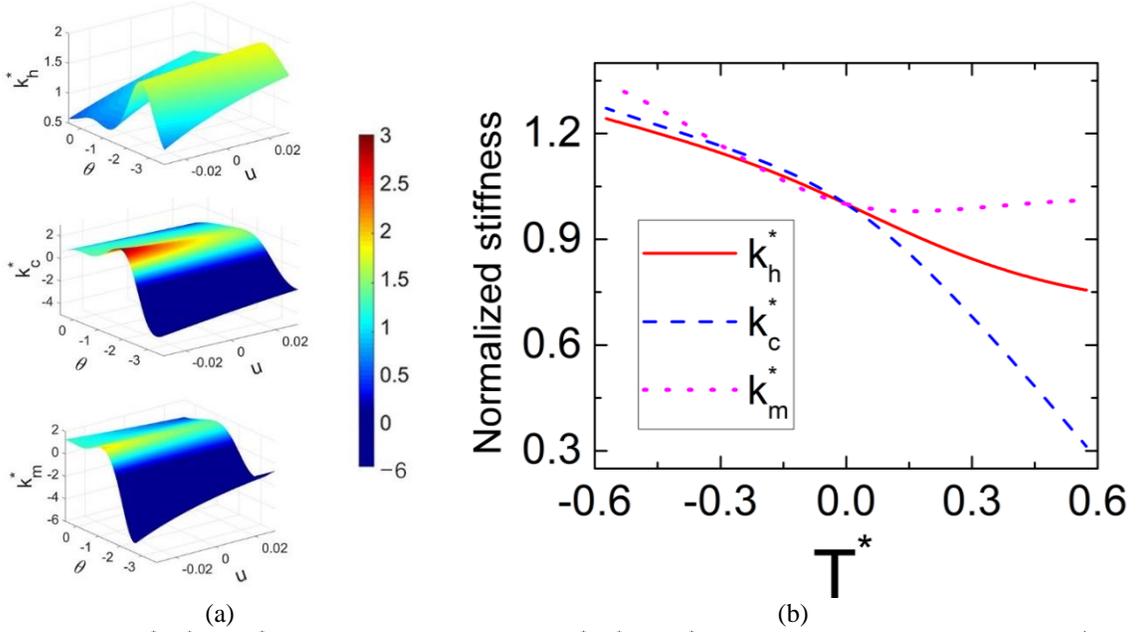

(a) (b)
**Fig. 2.** (a) $k_h^*$, $k_c^*$ and $k_m^*$ as functions of $u$ and $\theta$; (b) $k_h^*$, $k_c^*$ and $k_m^*$ change with varying static torque $T^*$.

For linear elastic wave propagations in a 1D infinite chain consisting of repetitive PTSs, the governing equation of the $n^{\text{th}}$ PTS cell can be expressed by using the obtained effective stiffness as

$$m\ddot{u}_n = k_h(u_{n+1}+u_{n-1}-2u_n)+k_c(\theta_{n+1}+\theta_{n-1}-2\theta_n)$$
$$J\ddot{\theta}_n = k_c(u_{n+1}+u_{n-1}-2u_n)+k_m(\theta_{n+1}+\theta_{n-1}-2\theta_n)$$
(7)

where $m$ and $J$ are the mass and moment of inertia of the PTS's disk, respectively. Considering harmonic wave excitations with angular frequency $\omega$ and periodic condition on the unit cell's boundaries, the governing equation can be rewritten as

$$\omega^2 \begin{bmatrix} m & 0 \\ 0 & J \end{bmatrix} \begin{Bmatrix} C_1 \\ C_2 \end{Bmatrix} = \begin{bmatrix} k_h \left( e^{ikh_0} + e^{-ikh_0} - 2 \right) & k_c \left( e^{ikh_0} + e^{-ikh_0} - 2 \right) \\ k_c \left( e^{ikh_0} + e^{-ikh_0} - 2 \right) & k_m \left( e^{ikh_0} + e^{-ikh_0} - 2 \right) \end{bmatrix} \begin{Bmatrix} C_1 \\ C_2 \end{Bmatrix} \qquad (8)$$

where $C_1$ and $C_2$ are the amplitudes of the axial and rotational displacements in the first PTS unit cell, respectively. An eigenvalue problem is then formed from Eq. (8) and the dispersion results for two wave modes can be calculated. By applying different static torques to the PTS, varying dispersion curves for torsional wave (rotational motion-dominated) and axial wave (axial motion-dominated) are shown in Fig. 3a and b, respectively. In the figures, the normalized frequency is defined as $f^* = 2\pi f h_0 / (k_{h_0}/m)^{1/2}$. The black lines are the dispersion curves when the chain is under zero external torque while the red lines and blue lines represent the dispersion curves when the chain is under positive (count-clockwise) and negative (clockwise) static torques. The inserted sketches in Fig. 3a and 3b demonstrate the mode shapes of the torsional and axial waves, respectively. For the torsional wave, both positive and negative torques increase the wave velocity due to the stiffening elastic response of each PTS cell (pink dot curve in Fig. 2b). However, for the axial wave, the positive torque decreases the wave velocity due to the softening response of the PTS cell while the negative torque increases the wave velocity due to the stiffening response of the PTS cell (red solid curve in Fig. 2b)

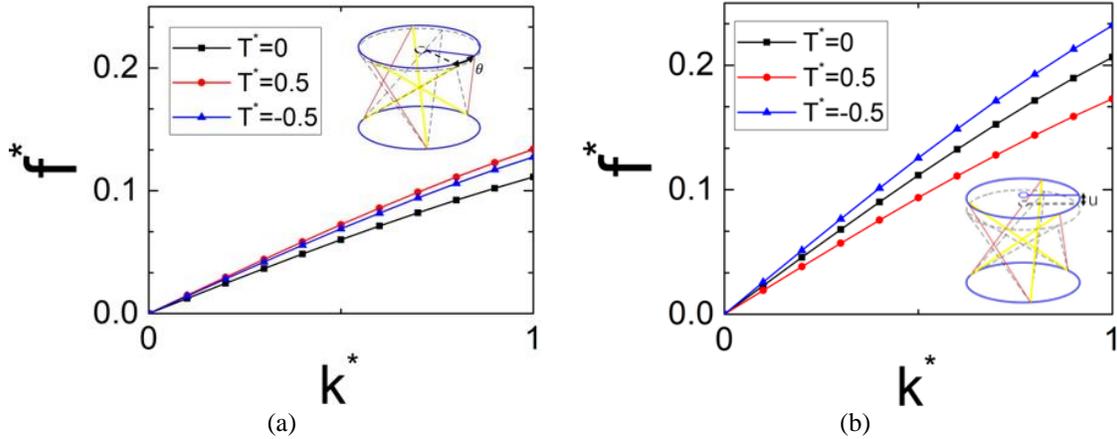

(a)　　　　　　　　　　　　　　(b)

**Fig. 3.** (a) The dispersion curves of torsional wave with different static torques; (b) the dispersion curves of axial wave with different static torques.

## 3. Designs of lightweight tensegrity metastructures

In this section, various lightweight tensegrity metastructures are designed based on the Bragg scattering mechanism as well as the local resonance mechanism to create desired bandgaps at targeted frequency ranges. Fig. 4a shows a Bragg-type tensegrity metastructure consisting of PTSs with alternating thick-hollow and thin-solid end disks. A unit cell of the tensegrity metastructure is selected inside the gray dashed rectangular region with a solid disk ($m$, $J_A$), a hollow disk ($m$, $J_B$) and two sets of bars and strings that connect the neighboring disks with the same chirality (marked in red color), where $J_A = J$ and $J_B = 1.81J$.

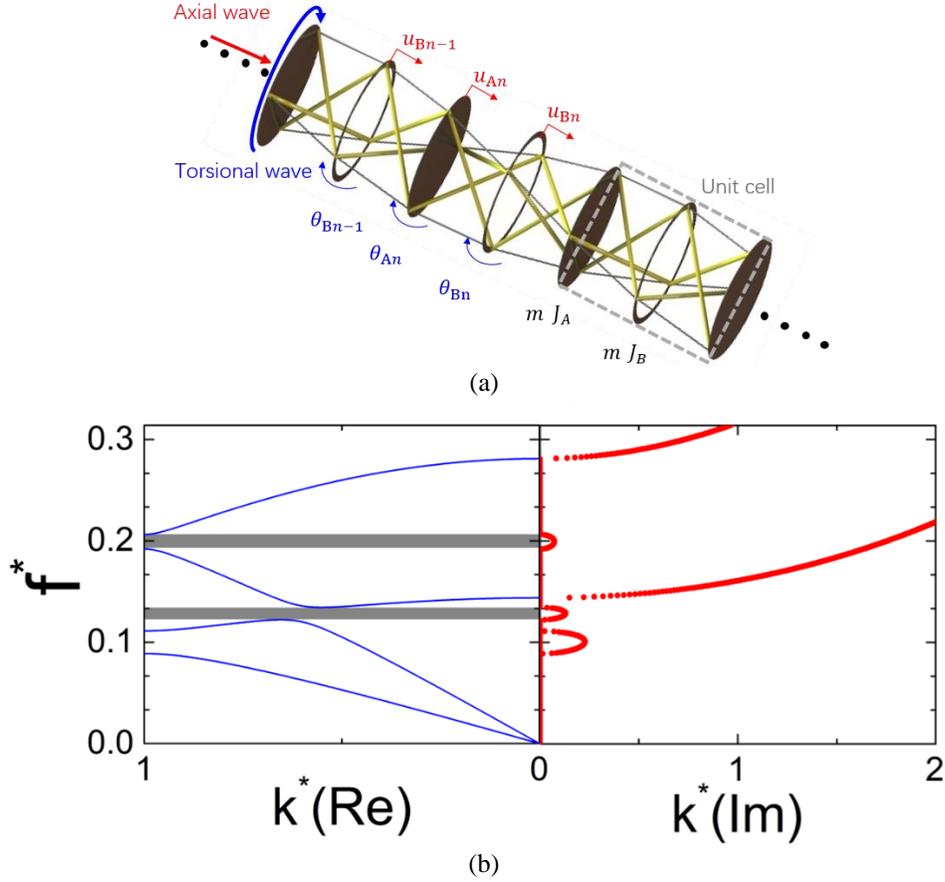

(a)

(b)

**Fig. 4.** (a) Schematic of the Bragg-type tensegrity metastructure with same chirality; (b) dispersion curves with real and imaginary wavenumbers. Gray-shaded regions indicate full-wave bandgaps.

Similar to Eq. (7), applying the harmonic wave assumption and Bloch theorem, the dispersion curves of the constructed tensegrity metastructure can be calculated by solving the eigenvalue problem obtained from the following governing equations.

$$\omega^2 \begin{bmatrix} m_A & 0 & 0 & 0 \\ 0 & J_A & 0 & 0 \\ 0 & 0 & m_B & 0 \\ 0 & 0 & 0 & J_B \end{bmatrix} \begin{Bmatrix} C_{A1} \\ C_{A2} \\ C_{B1} \\ C_{B2} \end{Bmatrix}$$

$$= \begin{bmatrix} k_{h1}+k_{h2} & k_{c1}+k_{c2} & -k_{h1}e^{-ikh_0}-k_{h2}e^{ikh_0} & -k_{c1}e^{-ikh_0}-k_{c2}e^{ikh_0} \\ k_{c1}+k_{c2} & k_{m1}+k_{m2} & -k_{c1}e^{-ikh_0}-k_{c2}e^{ikh_0} & -k_{m1}e^{-ikh_0}-k_{m2}e^{ikh_0} \\ -k_{h1}e^{ikh_0}-k_{h2}e^{-ikh_0} & -k_{c1}e^{ikh_0}-k_{c2}e^{-ikh_0} & k_{h1}+k_{h2} & k_{c1}+k_{c2} \\ -k_{c1}e^{ikh_0}-k_{c2}e^{-ikh_0} & -k_{m1}e^{ikh_0}-k_{m2}e^{-ikh_0} & k_{c1}+k_{c2} & k_{m1}+k_{m2} \end{bmatrix} \begin{Bmatrix} C_{A1} \\ C_{A2} \\ C_{B1} \\ C_{B2} \end{Bmatrix} \quad (9)$$

where $C_{A1}$, $C_{A2}$, $C_{B1}$, $C_{B2}$ are amplitudes of the axial and rotational displacements of the solid and hollow disks in the first unit cell, respectively. Fig. 4b shows the dispersion curves of the metastructure where both real and imaginary parts of the normalized wavenumber are calculated for each frequency point. First, two full-wave bandgaps (neither torsional wave nor axial wave can propagate) are found in the gray-shaded regions with zero real wavenumbers and non-zero imaginary wavenumbers. The central-maximum curves of $k^*$(Im) in both bandgaps indicate their Bragg-scattering origin (Zhu, et al., 2012; Wang, et al., 2004; Yu, et al., 2008). Second, another central-maximum curve of $k^*$(Im) is found in the frequency range just below the first full-wave bandgap where only propagating axial wave exists, which indicates a torsional wave bandgap generated by Bragg scattering. Third, two monotonically increasing curves of $k^*$(Im) can be found above $f^* = 0.281$ and $f^* = 0.144$ which are the two cutoff frequencies for the axial and torsional waves, respectively.

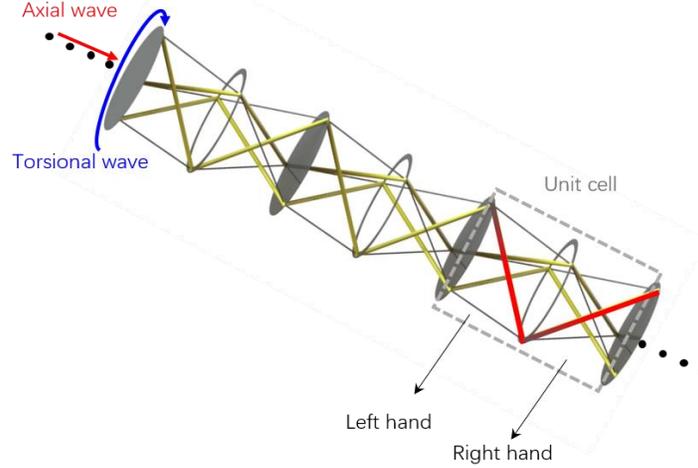

(a)

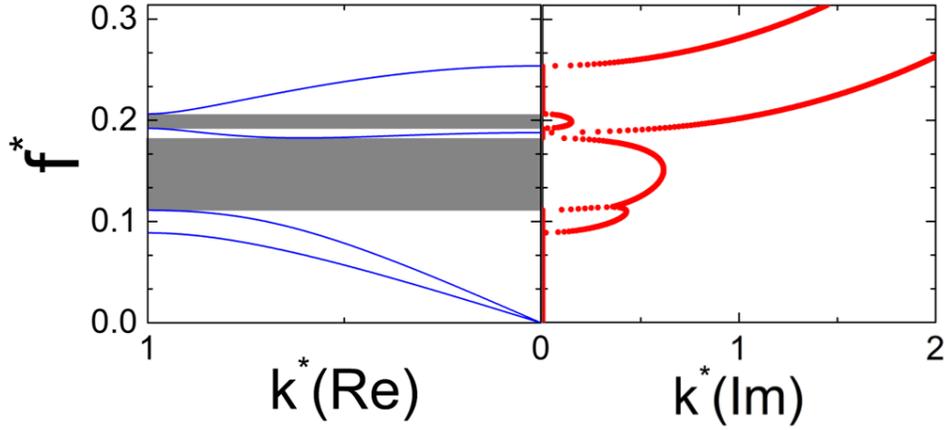

(b)

**Fig. 5.** (a) Schematic of the Bragg-type tensegrity metastructure with opposite chirality; (b) dispersion curves with real and imaginary wavenumbers. Gray-shaded regions indicate full-wave bandgaps.

Fig. 5a shows the schematic of a Bragg-type tensegrity metastructure with opposite chirality in each building unit cells. With oppositely chiral arrangements of the bars and strings (marked in red color) used for the two PTSs in each metastructure unit cell, different effective stiffness matrixes are applied to the governing equations. Fig. 5b shows the calculated dispersion curves of the new metastructure. First, it can be found that the first full-wave bandgap is greatly expended. The unchanged central-maximum profile of the $k^*$(Im) curve suggests the Bragg-scattering mechanism behind the enlarged bandgap. Second, the much larger $k^*$(Im) values in the first bandgap indicate stronger attenuations for both axial and torsional waves. Since neither the weight nor the size of the tensegrity metastructure increases, the introduced opposite chirality proofs to be a practical way to efficiently attenuate elastic wave propagations in a broad frequency range.

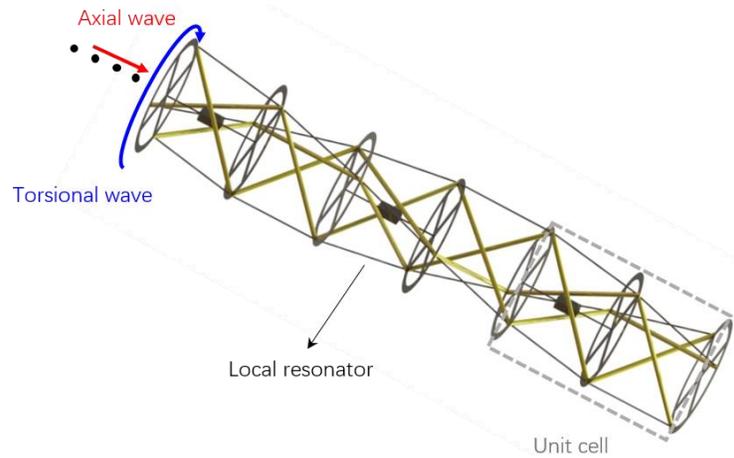

(a)

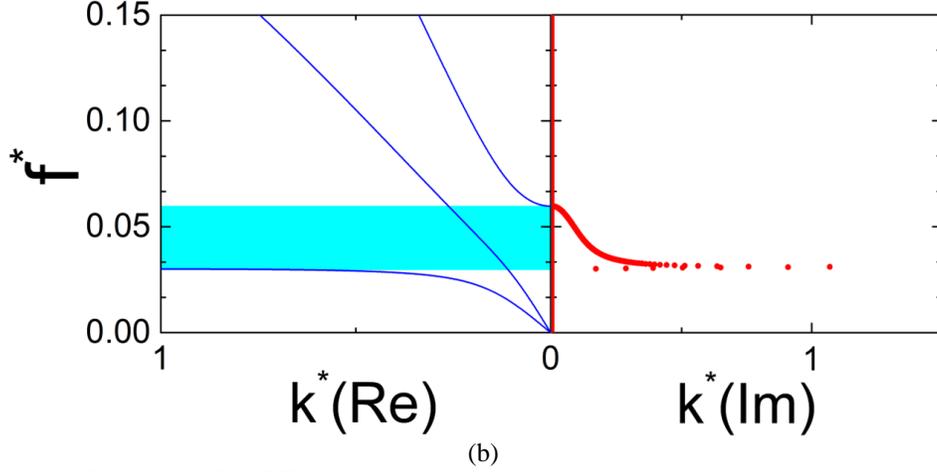

(b)

**Fig. 6.** (a) Schematic of the LR-type tensegrity metastructure; (b) dispersion curves with real and imaginary wavenumbers. Light blue-shaded regions indicate torsional wave bandgap.

Fig. 6a shows the schematic of a local resonance-based tensegrity metastructure. The introduced resonator in each unit cell consists of a cylindrical resonator ($m_r = 1.476m$) and two thin strings ($k_r = 0.22\ k_s$) that connects the resonator to the PTS's end disks. The small stiffness of the thin strings is chosen to ensure that the local resonance happens at a low frequency point while the diameter of the resonator is designed to fit inside the central space of the chiral bars and strings without blocking their motions. Cross-patterned end disks ($m_C$, $J_C$) with $m_C = 0.262m$ and $J_C = 0.422J_A$ are used in each PTS to compensate the added mass of the resonator and ensure that the overall weight of the metastructure keeps. Fig. 6b shows the dispersion curves of the LR-type metastructure. First, no full-wave bandgap can be found in the low frequency region. Second, an axial wave bandgap can be found in the light blue shaded region. The local resonance origin of the axial wave bandgap is evidenced by the lower edge-maximum profile of the $k^*$(Im) curve in the corresponding frequency range (Xiao, et al., 2011; Xiao, et al., 2012). Therefore, the LR-based tensegrity metastructure can block axial wave in a frequency range about four times lower than Bragg-based tensegrity metastructure while keep the size and weight of the entire structure unchanged.

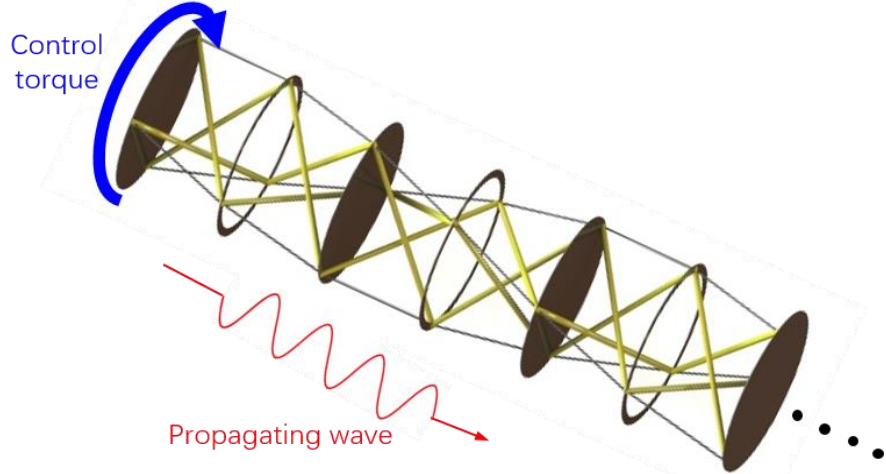

**Fig. 7.** Schematic of small-on-large tunability in the tensegrity metastructure.

In the previous section, quantitative analysis on the PTS's effective stiffness controlled by an external torque loading has been conducted statically (Fig. 2b) and dynamically (Fig. 3a and 3b). The following question arises naturally: can the band structure of the tensegrity metastructure be controlled by an external loading? The key to answer this question lies on the recently demonstrated 'small-on-large' tunability (Wang, et al., 2013; Wang, et al., 2014). The 'large' geometrically nonlinear deformation induced by a static control torque changes the effective stiffness of each PTS, the fundamental building block of the tensegrity metastructure, and therefore, affects the 'small' amplitude linear elastic wave propagation inside the metastructure, as shown in Fig. 7. Both positive and negative control torques can be applied to the metastructure and their effects on the dispersion curves are shown in Fig. 8a and 8b, respectively. By imposing a positive control torque, the lowest bandgap rises to a higher frequency region while the cutoff frequency drops, as shown in Fig. 8a. More interestingly, the second bandgap closes at around $f^*$=0.2 showing that the control torque can also be used to turn on/off the bandgap of the tensegrity metastructure (proof in the finite metastructure will be demonstrated in the next section). In the negative control torque case, not only the lowest bandgap but also the second bandgap and the cut-off frequency rise to higher frequency regions.

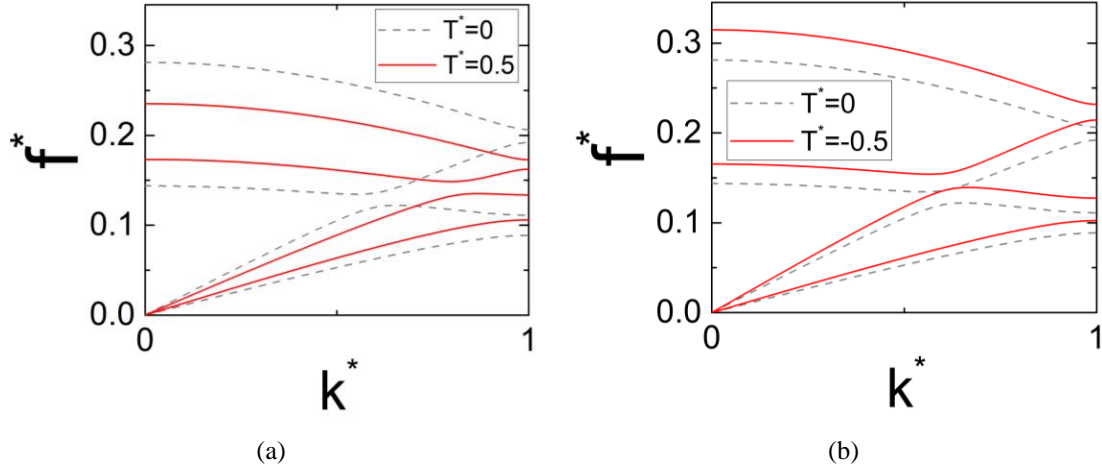

(a)                  (b)

**Fig. 8.** Dispersion curves of the tensegrity metastructure under (a) positive control torque and (b) negative control torque.

The prestress in each string of the PTS cell also provides an alternative way to tune the dispersion curves of the tensegrity metastructure. The prestress control is feasible in real tensegrity structures with a hydraulic actuator (Kmet and Platko, 2014). Fig. 9 shows the dispersion curves of the tensegrity metastructure with different prestresses in the strings. It can be found that decreasing the prestress barely changes the dispersion curves at the high frequency region. But it can compress dispersion curves below $f^*$=0.15 and move the two full-wave bandgaps to lower frequency regions.

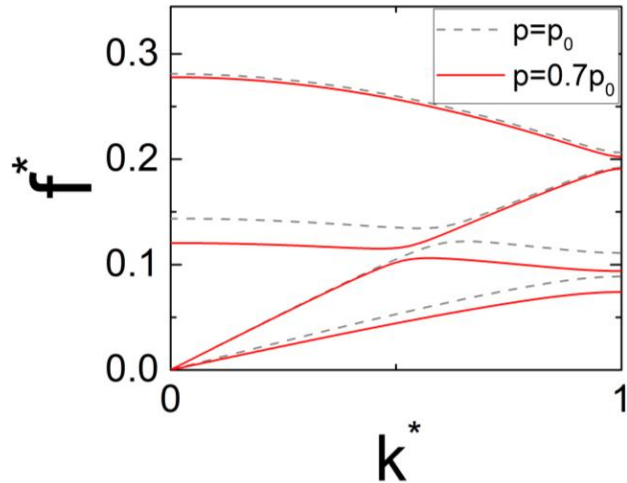

**Fig. 9.** Dispersion curves with different prestresses in the strings.

## 4. Vibration isolation in finite tensegrity metastructures

In the previous sections, wave propagations in the infinite tensegrity metastructures based on Bragg-scattering mechanism as well as local resonance mechanism have been systematically investigated. However, infinite structures are not realistic in the real-world engineering applications and therefore, vibration tests in the finite structures with defined boundary conditions should be conducted to characterize and validate the dynamic properties of the proposed tensegrity metastructures. The top parts in Figs. 10a, 10b and 10c show the schematics of the finite tensegrity metastructures with Bragg-type same-chirality unit cells, Bragg-type opposite-chirality unit cells and LR-type unit cells, respectively. Only side views are shown in the schematics with the blue thin lines, black thick lines and the green thin lines being the strings connecting PTSs, the elastic bars and the thin strings connecting local resonators, respectively. The solid, hollow and cross-linked disks are presented by the gray solid rectangles, the dash-line hollow rectangles and I-pattern rectangles, respectively. 20 unit cells are used in all three tensegrity metastructures. Harmonic axial and torsional force excitations are applied to one side of the metastructure and the other side is fixed. A frequency sweep is conducted in the normalized frequency range $f^* = (0, 0.34)$. Frequency-response functions (FRFs) then are defined for the axial and torsional waves as $FRF_C$ and $FRF_T$, respectively. For steady-state vibrations, the FRFs for the two wave modes can be defined as follows:

$$
\begin{aligned}
FRF_C &= 10\log(\frac{u_{20}}{u_1}) \\
FRF_T &= 10\log(\frac{\theta_{20}}{\theta_1})
\end{aligned}
\tag{10}
$$

where

$$u_j = \tilde{u}_j e^{-i\omega t}, j = 1 \text{ or } 20$$

$$\theta_j = \tilde{\theta}_j e^{-i\omega t}, j = 1 \text{ or } 20$$

are the axial and torsional displacements measured at the sensor points located at first end disks of the 1$^{st}$ and 20$^{th}$ metastructure unit cells.

The bottom parts of Fig. 10 show the $FRF_C$ and $FRF_T$ results for the three finite tensegrity metastructures. The gray-shaded and light blue-shaded zones represent the full-wave bandgap and the axial wave bandgap obtained from the infinite metastructure analysis, respectively. A vibration attenuation threshold is defined at -10dB for both $FRF_C$ and $FRF_T$ to identify the frequency range of the attenuation zone in the finite metastructure (Zhu, et al., 2014). In Fig. 10a, two normalized frequency ranges $f^* = (0.122, 0.134)$ and $(0.192, 0.206)$ are found to have both $FRF_C$ and $FRF_T$ below -10dB, which almost coincide with the two gray-shaded zones and therefore, validate the predicted full-wave bandgaps in the infinite metastructures. Apparently low amplitude of the $FRF_T$ curve can be found in a small frequency range just below the first gray-shaded zone as well as the frequency range above $f^* = 0.144$, which is due to the Bragg-scattering-induced torsional wave bandgap and the cutoff frequency of the torsional wave at $f^* = 0.144$. It should be noticed that the axial waves with small coupled torsional motions are not affected in these two frequency ranges and still contributes to the $FRF_T$ whose amplitude is therefore, above -10dB. After the cutoff frequency at $f^* = 0.281$, neither $FRF_T$ nor $FRF_C$ is above -10dB. In Fig. 10b, the normalized frequency ranges $f^* = (0.111, 0.182)$ and $(0.192, 0.206)$ are found to have both $FRF_C$ and $FRF_T$ below -10dB, which also coincide well with two gray-shaded full-wave bandgaps in the infinite opposite-chirality tensegrity metastructure. Both $FRF_T$ and $FRF_C$ are close to -120dB at the central frequency point in the first attenuation zone, which can be well explained by the much larger $k^*(Im)$ value at the same frequency point in Fig. 5b. Also, low-amplitude $FRF_T$ curves can be found in the torsional wave bandgap and above the torsional wave cutoff frequency at $f^* = 0.21$. Finally, significantly decreased $FRF_C$ curve can be found in the frequency range $f^* = (0.03, 0.06)$ in Fig. 10c, which is in good agreement with the local resonance-induced axial wave bandgap (light blue-shaded zone) found in the corresponding infinite LR-type metastructure. In this frequency range, the $FRF_C$ amplitudes at most frequency points are below -10dB except for those at the three peaks which are generated by the global resonant motions of the entire finite tensegrity metastructure.

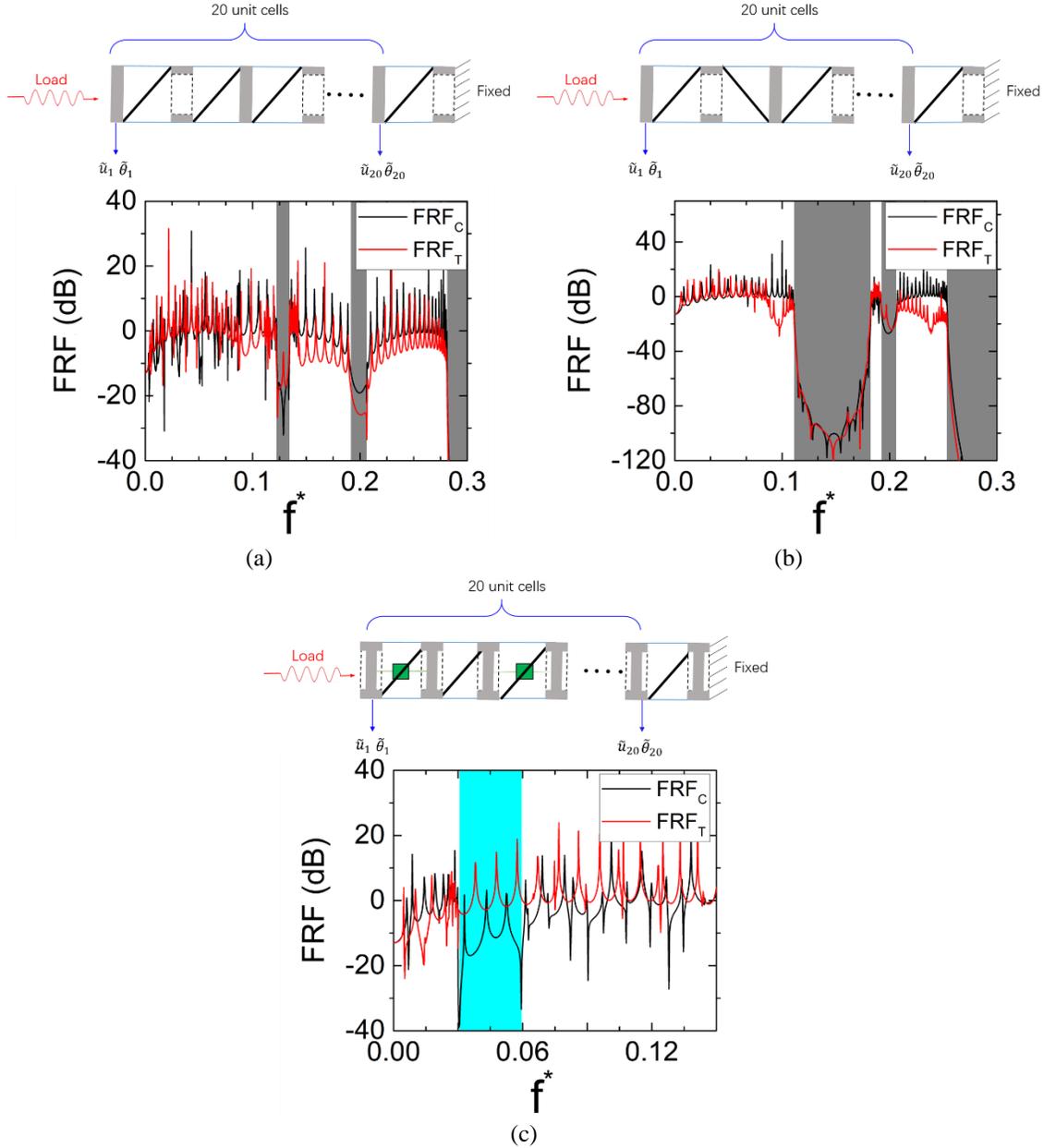

**Fig. 10.** (a) Schemetic and FRF results of the finite tensegrity metastructure with 20 Bragg-type same-chirality unit cells; (b) schematic and FRF results of the finite tensegrity metastructure with 20 Bragg-type opposite-chirality unit cells; (c) schematic and FRF results of the finite tensegrity metastructure with 20 LR-type unit cells.

In order to validate the 'small-on-large' tunability in the proposed tensegrity metastructure, Fig. 11 shows the FRF results of finite tensegrity metastructures with Bragg-type same-chirality unit cells under different external control torques. In the figure, the gray dash lines, the red solid lines and the blue dash-dot lines are the FRFs with zero torque ($T^*=0$), positive torque ($T^*=0.5$) and negative torque ($T^*=-0.5$), respectively. Fig.11a and 11b demonstrate the result of the axial wave ($FRF_C$) and torsional wave ($FRF_T$), respectively. First, it can be found that the first valleys of the $FRF_C$ and $FRF_T$ curves (FRFs≤10dB) move from the frequency range $f^*=$ (0.111,0.182) at $T^*=0$ to higher frequency ranges $f^*=$ (0.135,0.148) and $f^*=$ (0.139,0.154) at $T^*=0.5$ and $T^*=-0.5$, respectively. Same trend and almost overlapped frequency ranges of full-wave bandgaps can be found in the dispersion curves of the infinite metastructure under the same control torques, as shown in Fig. 8, which successfully validate the tunability of finite metastructures. The lonely peaks in the $FRF_C$ valleys are due to the finite structure resonant motions. Second, it is noticed that no second FRF valley can be found for $T^*=0.5$, which indicates that the on/off bandgap switch ability can also be found in the finite metastructure. Finally, three cut-off frequencies can be found in the FRFs results at the frequency points $f^*=$ 0.281, $f^*=$ 0.235 and $f^*=$ 0.315 for the $T^*=0$, $T^*=0.5$ and $T^*=-0.5$, respectively. Comparing with the cut-off frequencies predicted in Fig. 8, very good agreements can be observed.

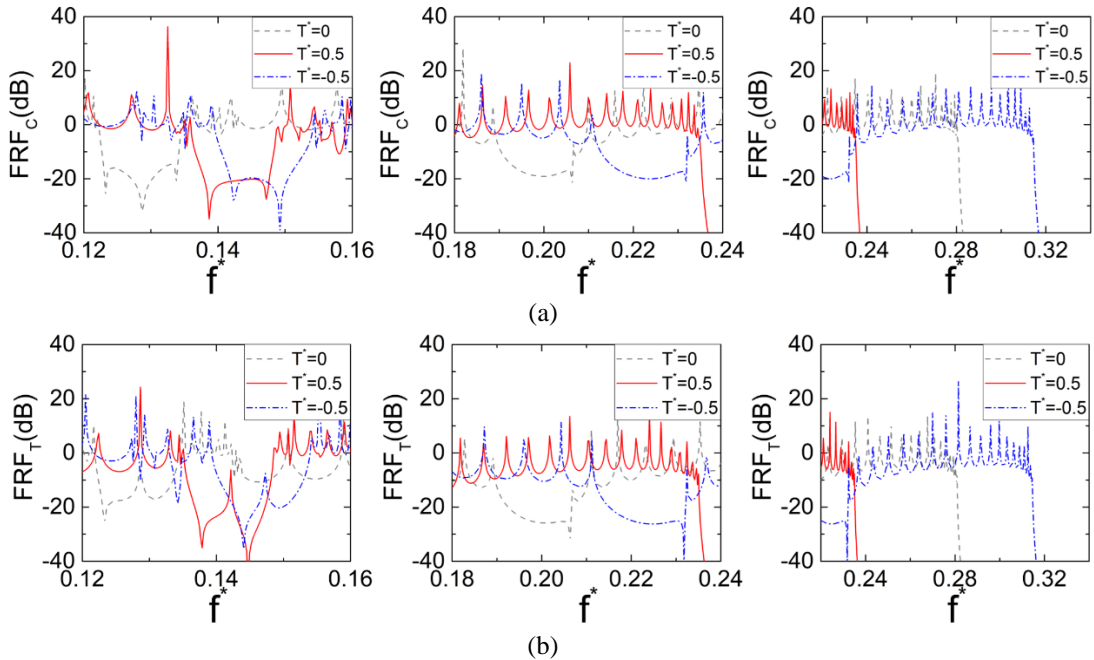

**Fig. 11.** (a) $FRF_C$ results of the finite tensegrity metastructure with 20 Bragg-type same-chirality unit cells under different control torque loadings; (b) $FRF_T$ results of the finite tensegrity metastructure with 20 Bragg-type same-chirality unit cells under different control torque loadings.

## 5. Conclusions

A theoretical model with coupled axial-torsional effective stiffness matrix is developed to study the band structures of metastructures consisting of prismatic tensegrity cells. Unit cell designs are conducted based on both Bragg scattering and local resonance mechanism to build tensegrity metastructures with bandgaps at desired frequency ranges. It is noticed that unit cell with opposite chirality can lead to broadband attenuations for both axial and torsional waves. Moreover, tunable wave propagations are investigated by two approaches: (i) harnessing the geometrically nonlinear deformation of the periodical tensegrity prisms under global control torques to achieve 'small-on-large' tunability; (ii) adjusting the prestress in the tensegrity strings with active components to achieve fine adjustment of the band structure. Finally, frequency responses studies on the finite structures are preformed to validate the wave attenuation ability as well as the tunability of the proposed tensegrity metastructures.

# Acknowledgements

The authors are grateful for the financial support from the National Natural Science Foundation of China (Nos. 11521062, 11632003, and 11290153). R. Zhu acknowledges the support from the Thousand Young Talents Program of China.

## Appendix

The reference configuration of the PTS is shown in Fig. 1b, where the radius of the end-disks, the height of the PTS and the relative angle of the two end-disks are $R$, $h$ and $\phi_0$, respectively. In this study, we assume that the two end-disks are maintained to be parallel and the central axis of the PTS, OO', is always along the z direction. Therefore, only two DOFs of the PTS are permitted, which are the relative rotational angle and the relative axial displacement between the two end-disks.

If the bottom disk is fixed, the top disk of the PTS has coupled axial and torsional motions. The coupled axial displacement and rotational angle can be defined as $u$ and $\theta$, respectively. It is noticed that the value of $\theta$ should be constrained within $-7\pi/6$ and $\pi/6$ to avoid the touch of the strings and the touch of the bars, respectively (Oppenheim and Williams, 2000). Then, the global coordinates of the three joints in Fig. 1b can be expressed as:

Joint A: $x_A = R$, $y_A = 0$, $z_A = 0$
Joint B: $x_B = R\cos(2/3\pi)$, $y_B = R\sin(2/3\pi)$, $z_B = 0$  (A.1)
Joint A': $x_{A'} = R\cos(\phi_0 + \theta)$, $y_{A'} = R\sin(\phi_0 + \theta)$, $z_{A'} = h+u$

The current length of the bars and strings can be given based on the coordinates of the joints as

$$L_b = \sqrt{[R\cos(\phi_0 + \theta) - R]^2 + [R\sin(\phi_0 + \theta)]^2 + (h+u)^2}$$
$$L_s = \sqrt{[R\cos(\phi_0 + \theta) - R\cos(2/3\pi)]^2 + [R\sin(\phi_0 + \theta) - R\sin(2/3\pi)]^2 + (h+u)^2}$$
(A.2)

The prestresses in the bars and strings can be calculated as :

$$p_b = k_b(L_b^0 - L_b^n)$$
$$p_s = k_s(L_s^0 - L_s^n)$$
(A.3)

where $L_b^0 = L_b(u=0, \theta=0)$ and $L_s^0 = L_s(u=0, \theta=0)$; $k_b$ is the stiffness of the bar and $k_s$ is the stiffness of the string; $L_b^n$ and $L_s^n$ are the nature length of the bar and the string, respectively. The potential energy of the tensegrity structure can also be calculated as:

$$E = \frac{3}{2}k_b \Delta L_b^2 + \frac{3}{2}k_s \Delta L_s^2 \qquad (A.4)$$

where $\Delta L_b = L_b - L_b^n$ and $\Delta L_s = L_s - L_b^n$ are the length changes of the bars and strings, respectively.

For the tensegrity structure under external axial force and torque, the total potential energy of the system, $\Pi$, is the sum of structure strain energy and the work down by external force and torque as:

$$\Pi = E - T\theta - Fu \qquad (A.5)$$

where $T$ denotes the externally applied torque, $F$ denotes the externally applied force. The principle of minimum potential energy can be used to determine the equilibrium of the system under load. Thus, force and torque on the PTS can be expressed by the rotating angle and the axial displacement, the detailed expressions are given as follow:

$$F = \frac{\partial \Pi}{\partial u} = 3(h_0 + u)[k_b + k_s - \frac{k_b L_b^n}{\sqrt{h_0^2 + 2R^2 + 2h_0 u + u^2 - 2R^2 \cos(\theta)}}$$
$$- \frac{k_s L_s^n}{\sqrt{h_0^2 + 2R^2 + 2h_0 u + u^2 - R^2 \cos(\theta) - \sqrt{3}R^2 \sin(\theta)}}] \qquad (A.6)$$

$$T = \frac{d\Pi}{d\theta} = 3R^2[k_b \sin(\theta) - k_s \cos(\frac{\pi}{6} - \theta) - \frac{L_b^n \sin(\theta)}{\sqrt{h_0^2 + 2R^2 + 2h_0 u + u^2 - 2R^2 \cos(\theta)}}$$
$$+ \frac{k_s L_s^n \cos(\frac{\pi}{6} - \theta)}{\sqrt{h_0^2 + 2R^2 + 2h_0 u + u^2 + R^2 \cos(\theta) - \sqrt{3}R^2 \sin(\theta)}}] \qquad (A.7)$$

Their first order Taylor expansion can be obtained as $F = k_h u + k_c \theta$ and $T = k_c u + k_m \theta$ with detailed efficient stiffness expressions as follow:

$$k_h = \frac{\partial F}{\partial u} = \frac{3k_b\{-2L_b^n R^2 + 2L_b^n R^2 \cos(\theta) + [h_0^2 + 2R^2 + 2h_0 u + u^2 - 2R^2 \cos(\theta)]^{3/2}\}}{[h_0^2 + 2R^2 + 2h_0 u + u^2 - 2R^2 \cos(\theta)]^{3/2}}$$
$$+ \frac{3k_s[-2L_s^n R^2 - L_s^n R^2 \cos(\theta) + \sqrt{3}L_s^n R^2 \sin(\theta) + (h_0^2 + 2R^2 + 2h_0 u + u^2 + R^2 \cos(\theta) - \sqrt{3}R^2 \sin(\theta))^{3/2}]}{[h_0^2 + 2R^2 + 2h_0 u + u^2 + R^2 \cos(\theta) - \sqrt{3}R^2 \sin(\theta)]^{3/2}} \qquad (A.8)$$

$$k_c = \frac{\partial F}{\partial \theta} = \frac{\partial T}{\partial u} = \frac{-3k_s(h_0+u)(R^2\cos(\pi/6-\theta))}{[(h_0+u)^2+R^2(2+\cos(\theta)-\sqrt{3}\sin(\theta))]}$$
$$+\frac{3k_s R^2(h_0+u)\cos(\pi/6-\theta)(-L_s^n+\sqrt{h_0^2+2R^2+2h_0u+u^2+R^2\cos(\theta)-\sqrt{3}R^2\sin(\theta)})}{[h_0^2+2R^2+2h_0u+u^2+R^2\cos(\theta)-\sqrt{3}R^2\sin(\theta)]^{3/2}}$$
$$+\frac{3k_b L_b^n R^2(h_0+u)\sin(\theta)}{[h_0^2+2R^2+2h_0u+u^2-2R^2\cos(\theta)]^{3/2}}$$

(A.9)

$$k_m = \frac{\partial T}{\partial \theta} = 3R^2\cos(\theta)(k_b - \frac{k_b L_b^n}{\sqrt{h_0+2R^2+2h_0u+u^2-2R^2\cos(\theta)}})$$
$$+\frac{3R^4 k_b L_b^n \sin^2(\theta)}{[h_0^2+2R^2+2h_0u+u^2-2R^2\cos(\theta)]^{3/2}} - 3R^2 k_s \sin(\pi/6-\theta)$$
$$+\frac{3R^4 k_s L_s^n \cos^2(\pi/6-\theta)}{[h_0^2+2R^2+2h_0u+u^2+R^2\cos(\theta)-\sqrt{3}R^2\sin(\theta)]^{3/2}}$$
$$+\frac{3R^2 k_s L_s^n \sin(\pi/6-\theta)}{\sqrt{h_0^2+2R^2+2h_0u+u^2+R^2\cos(\theta)-\sqrt{3}R^2\sin(\theta)}}$$

(A.10)